# Programmable graphene doping via electron beam irradiation

Yangbo Zhou[1], Jakub Jadwiszczak[1], Darragh Keane[2], Ying Chen[3], Dapeng Yu[4],

Hongzhou Zhang[1]

[1]School of Physics, CRANN and AMBER, Trinity College Dublin, Dublin 2, Ireland

[2]School of Chemistry, CRANN and AMBER, Trinity College Dublin, Dublin 2, Ireland

[3]Institute for Frontier Materials, Deakin University, Waurn Ponds, VIC 3216, Australia

[4]School of Physics, Peking University, Beijing, 100871, China

Correspondence and request for materials should be addressed to Hongzhou Zhang (hozhang@tcd.ie)

**Abstract:**

   Graphene is a promising candidate to succeed silicon based devices and doping holds the key to graphene electronics. Conventional doping methods through surface functionalization or lattice modification are effective in tuning carrier densities. These processes, however, lead to degradation of device performance because of structural defect creation. A challenge remains to controllably dope graphene while preserving its superlative properties. Here we show a novel method for tunable and erasable doping of on-chip graphene, realized by using a focused electron beam. Our results demonstrate site-specific control of carrier type and concentration achievable by modulating the charge distribution in the dielectric substrate. Thereby, the structural integrity and electrical performance of graphene are preserved, and the doping states are rewritable. Different logic operations were thus implemented in a single graphene sheet. By extending this method to other two-dimensional materials, this work lays out a blueprint for powerful yet simple means of incorporating two-dimensional materials into prospective electronic technologies.

**Introduction**

Graphene is a zero gap semi-metal[1,2]. By tuning the Fermi level in the Dirac cone, both p-type and n-type doping states can be realised. This bipolar doping, with a carrier density up to $10^{14}$ cm$^{-2}$, has been achieved through surface functionalization[3-6] and lattice modification[7-11]. However, these methods inevitably introduce permanent defective scattering centres reducing the carrier mobility[4,12]. The resultant device thus has limited performance without any tunability in conductance. Electrostatic field effect tuning can also modulate the carriers via the application of a back[13] or top gate[14]. The doping profile can be reversibly and finely controlled by the electrostatic field while maintaining the high quality of graphene[15,16]. Logic gates from ambipolar graphene have been constructed by electrostatically controlled doping[17]. However, this involves fabrication of complex device architecture. A rapid and efficient mechanism for fine control of doping concentration in graphene has yet to be demonstrated. In order for new device architecture to be tested, prototyping with reusable samples is highly desirable and necessary for graphene to become ubiquitous in future electronic technologies.

Here, we report site-specific reversible bipolar graphene doping with high mobility. This is enabled by injecting charge into the supporting substrate with a focused scanning electron beam. The doping profile in graphene can be precisely modulated by the location and dose of electron irradiation and is entirely erasable. Hence, both n-type and p-type doping can be achieved in the same device. We put forward a quantitative model to explain the e-beam induced doping in graphene, which can be extended to molybdenum disulfide (MoS$_2$). The localized and erasable nature of this technique enables the fabrication of multiple logic gates using the same graphene device. Regarding the advancement of two-dimensional (2D) electronics, this work not only provides a new insight into the nature of the field effect in graphene but also establishes a highly efficient way to prototype 2D logical devices for future electronic applications.

Results

**E-beam induced doping in graphene and its controllability**

By controlling the electron beam irradiation, the electrical transport properties of a graphene field-effect transistor (FET) can be modified (Fig. 1 and Supplementary Figs. 1-3). As shown in Figure 1a, such a beam was employed inside a scanning electron microscope (SEM), with the characterization of the consequent electrical transport taking place *in situ*. Gate response of the FET clearly demonstrates the doping phenomena induced by the e-beam irradiation. The pristine graphene shows an ambipolar gate response with the Dirac point near $-2\,V$ (black curve in Fig. 1b), indicating a low n-type carrier concentration of $\sim 1.5 \times 10^{11}\,cm^{-2}$ without gate biasing. This intrinsic low doping level is altered after e-beam irradiation, with the carrier type modulated by varying the e-beam energy. For example, after the graphene is exposed to a 2 keV e-beam, the gate response shifts to the left (red curve in Fig. 1b). The Dirac point moves from $-2\,V$ to $-55\,V$ after an irradiation dose of $10^{14}\,e^-/cm^2$ is applied, indicating strong n-type doping with a doping level of $\sim 4 \times 10^{12}\,cm^{-2}$. On the contrary, a 30 keV e-beam with a dose of $10^{15}\,e^-/cm^2$ shifts the Dirac point of the same graphene FET to $+11\,V$, producing hole doping. The zero-gate conductance of graphene increases with the increase of Dirac point shift, indicating the doping level. Figure 1c shows the zero-gate conductance as a function of beam energy which has a minimum close to that of the pristine graphene around 7 keV. Together with the gate responses, this means that we can alter the doping state in graphene by varying the kinetic energy of the incident electrons. A 7 keV beam corresponds to a non-doping case and marks the transition point from n-type doping at a lower beam energy to p-type at higher beam energies.

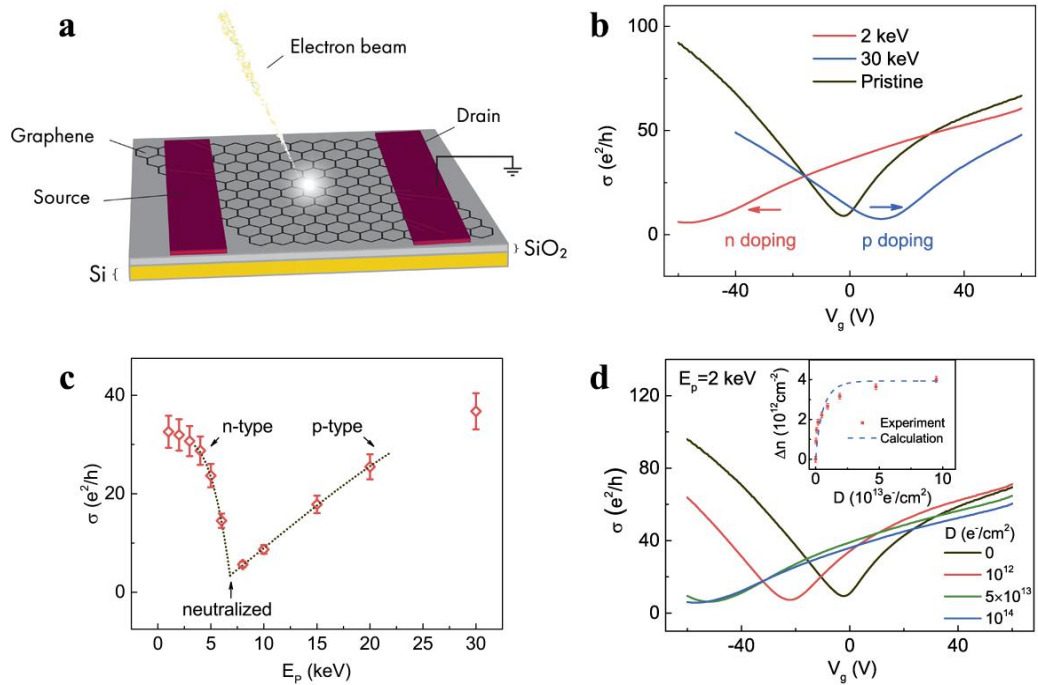

**Figure 1| Electron beam induced doping in graphene. a**, Sketch of the e-beam irradiation of a graphene device in a field effect geometry. **b**, Control of doping type by varying e-beam energies. **c**, Zero gate conductivity as a function of beam energy. The two dashed lines are the polynomial fittings to the data from 1- 6 keV and 8-20 keV respectively. **d**, Control of doping level with different irradiation doses using a 2 keV e-beam. Inset: evolution of carrier concentration with increasing dose; red squares are the experimental data, the blue dashed line is the theoretical calculation.

We can also control the carrier concentrations in the graphene flake (i.e. doping level) by tuning the electron dose. Figure 1d shows the gate response of the graphene FET under a 2 keV e-beam at different irradiation doses. A dose of $10^{12}\ e^-/cm^2$ shifts the Dirac point from $-2\ V$ to $-22\ V$ (red curve in Fig. 1d). As the irradiation dose increases, the doping level becomes higher and the Dirac point is shifted to $-50\ V$ after a dose of $5\times 10^{13}\ e^-/cm^2$. The carrier concentration increases rapidly under a low dose below $10^{13}\ e^-/cm^2$, but saturates to a value of $\sim 4\times 10^{12}\ cm^{-2}$ when the dose is increased up to $10^{14}\ e^-/cm^2$ as shown in the inset of Figure 1d. We measured six monolayer-graphene FETs in total and they all exhibited the doping behavior with a similar saturation level around $4\times 10^{12}\ cm^{-2}$ after the 2 keV e-beam irradiation (Supplementary Figs. S1 – S3).

**Stability and erasability of e-beam doping**

The n-type and p-type doping states exhibit different retention times. The high n-type doping level can be maintained in vacuum for a long period of time (over 16 hours) with a small decay rate ($\sim 1.9\times 10^7 cm^{-2}\cdot s^{-1}$, see Supplementary Fig.S4), while the p-type doping state decays much faster with decay rate $> 9\times 10^{10} cm^{-2}\cdot s^{-1}$ (estimated from Fig. 2c). The gate sweep also affects the doping states. As shown in Figure 2a, for the n-type doping induced by a 2 keV beam with a dose of $10^{13}\ e^-/cm^2$, the Dirac point slightly shifts a further ~ 2 V after 3 cycles of back gate sweep between -60 V and 60 V (solid lines), indicating the n-type doping is quite stable under the gate operation. On the contrary, for the p-type doping induced by a 30 keV e-beam with a dose of $10^{14}\ e^-/cm^2$, the Dirac point shifts from $\sim +8\ V$ to $-13\ V$ after 3 cycles of gate sweep (dashed lines in Fig. 2a).

The stable n-type doping can be erased by two methods. One is to expose the device to air. As shown in Figure 2b, the n-type states created by 2 keV e-beam irradiation (red curve) disappear after the device is exposed to air and the graphene exhibits weak p-type doping (blue curve) due to surface adsorbates[3]. After the graphene FET is placed back into vacuum (green curve), it shows a similar doping level comparable to its initial states prior to the irradiation (black dashed curve). This means that we can completely erase the beam-induced n-type doping and initialize the device to an undoped state by air exposure (Supplementary S6 for nitrogen). The other way to erase the n-type doping is to irradiate the device using higher beam energies. This is demonstrated by monitoring the graphene conductivity (with a floating back gate). As shown in Figure 2c, the graphene is initially n-doped with a conductivity of ~ 23 $e^2/h$, corresponding to an electron carrier density of ~ $4 \times 10^{12}\ cm^{-2}$. At $t = 20$ s, a 30 keV e-beam (beam current $I_b$ = 130 pA) starts to irradiate the graphene flake, and the conductivity quickly drops to a value close to that of the minimal conductivity at the charge neutral point, indicating the disappearance of the n-type doping. It then gradually recovers to ~ 18 $e^2/h$ over the next 20 seconds (red shadowed areas). This is due to the p-type doping with a hole density of ~$2 \times 10^{12}\ cm^{-2}$. As expected, the unstable p-type doping decays spontaneously after the e-beam is switched off. The subsequent irradiation with a 2 keV e-beam erases the p-type doping and the conductivity further reduces to the same minimal value before it starts to recover again. After the 2 keV e-beam is switched off, the conductivity maintains a constant level reflecting the stability of the n-type doping. This demonstrates that the doping states can be altered *in situ* by e-beam irradiation with appropriate beam energy, establishing the feasibility of a beam-writing cycle that flips the type of doping in graphene.

The increase of such writing cycles may degrade the device performance by cumulative beam damage and/or contamination[18,19]. We investigate the device performance with a high dose delivered by a 2 keV e-beam. The ambipolar gate response can be maintained and the beam-induced n-type doping is still observable under a dose up to $10^{17}\ e^-/cm^2$ (See Supplementary section 4). However, for the same doping dose of $10^{14}\ e^-/cm^2$, the doping level reduces from ~$5 \times 10^{12}\ cm^{-2}$ to ~$1.5 \times 10^{12}\ cm^{-2}$ as the cumulative dose increases from $10^{14}\ e^-/cm^2$ to $10^{17}\ e^-/cm^2$ (Fig. 2d). The ambipolar gate response and n-type doping in the graphene FET fail at a dose of $10^{18}$ $e^-/cm^2$. Therefore, we conclude that the doping is programmable up to $10^4$ cycles with a writing dose of $10^{13}$ $e^-/cm^2$ per cycle.

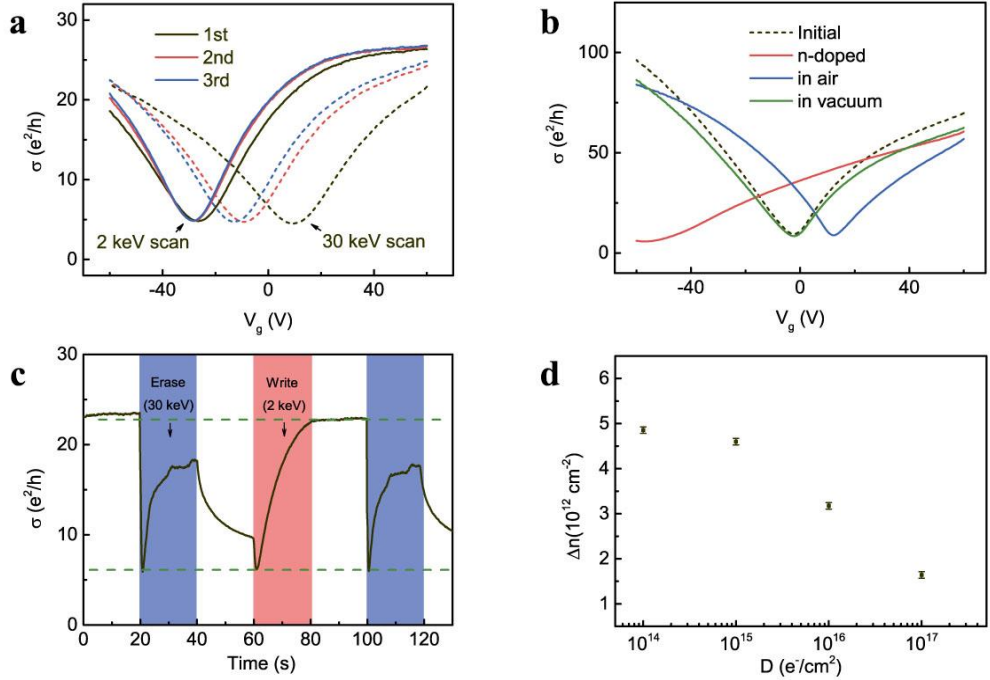

**Figure 2| Stability and erasability of e-beam induced doping. a,** Gate responses reveal the stability of doping states over three readings. **b,** Erasure of stable n-type doping by exposing to air. **c,** Switching between n-type and p-type doping by alternating beam energies. **d,** Reduction of doping behavior under large irradiation doses. The x-axis $D$ shows the cumulative dose, the y-axis $\Delta n$ is change of doping level induced by the writing dose of $10^{14}$ $e^-/cm^2$ after the graphene is irradiated by the cumulative dose $D$.

**Physical model of beam-induced graphene doping**

As illustrated in Figures 3a and 3b, the electron beam adjusts graphene doping by controlling the charge distribution in the substrate. The doping arises from a controlled balance between injection of primary beam, emission of secondary electrons and diffusion of charges in the substrate. The residual charges in the substrate generate an electric field, which shifts the Fermi level of the supported graphene and adjusts the type and density of carriers. This field effect is similar to that applied by a back gate, while the direct writing of the primary electron beam enables site-specific doping which will be demonstrated later. The charge concentration in the graphene is:

$$n_{gra} = -(\sigma - 1)De \quad (1)$$

where $\sigma$ is the yield of electron emission from the surface due to the primary beam irradiation, $D$ the beam dose (number of incident electrons per unit area) and $e$ the electron charge (Supplementary section 5). The doping type is determined by the sign of the charge concentration which can be toggled by tuning the emission yield $\sigma$. For example, a value of $\sigma$ greater than unity results in a negative carrier density in graphene, i.e. n-type doping.

The energy of the primary beam regulates the emission yield, and a unity $\sigma$ occurs at ~ 5-6 keV for the SiO$_2$/Si substrate[20,21]. When a beam of lower energies is applied, the emission yield is larger than 1, generating positive charges bound at the substrate surface inducing n-type doping in graphene (Fig. 3a). At higher beam energies the yield becomes lower than unity (e.g. < 0.2 at 30 keV), causing negative surface charging of the substrate and inducing p-type doping in graphene (Fig. 3b). This model provides quantitative explanation to the dependence of charge density in graphene on the irradiation dose (the blue dashed line in the inset of Fig. 1d, also see Supplementary section 5).

The range of the electron beam and the diffusion of the charges in the substrate control the stability of the doping. The electron range for a 2 keV beam into silicon dioxide is less than 100 nm, while it increases to 10 µm for a 30 keV e-beam (Supplementary Fig. S10). For the 2 keV e-beam irradiation, all injected charges remain in the SiO$_2$ layer (Fig. 3a) with a low drift mobility of $20\ cm^2/V \cdot s$ for electrons[22] (the holes are quite immobile[23] ~2 × $10^{-7} \ cm^2/V \cdot s$), resulting in the stable n-type doping. However, the majority of 30 keV primary electrons are injected into the highly doped Si substrate where the electron drift mobility ($\sim 100 \ cm^2/V \cdot s$) is about five times larger than that in the silicon oxide[24]. Consequently, the back gate bias will facilitate the spread of the electrons which attenuates the field effect and destabilizes the p-type state.

The erasing of doping states in the graphene device occurs through charge compensation in the substrate. For example, by exposing the device to air, all the excess surface charges will be quenched through charge transfer between adsorbed polar molecules such as water (Supplementary Fig. S6). This corroborates the surface charge model for the doping. Alternatively, the induced doping state can be erased by generating opposite charges in the substrate by varying the primary beam energy as demonstrated in Figure 2c.

These surface bound charges act as long-range scattering centers in graphene that affect its transport properties. According to the the semi-classical Boltzmann formalism[25-27], the conductivity of graphene is modulated by the carrier density (n) and the impurity density ($n_{imp}$),

$$\frac{1}{\sigma(n)} = \frac{1}{Ce}\left|\frac{n_{imp}}{n}\right| + \frac{1}{\sigma_{res}} \quad (2)$$

where $C = 5 \times 10^{15} V^{-1} s^{-1}$, a constant determined by the screened Coulomb potential[26,27] and $\sigma_{res}$ is the residual conductivity at the charge neutral point (Supplementary section 6). The carrier mobility can be extracted from experimental data by using Equation (2). Figure 3c shows the relationship between the mobility and the e-beam dose under 2 keV beam irradiation. The mobility is inversely proportional to the electron dose (inset of Fig. 3c). Both the electron and hole mobilities decrease from $\sim 10000 \ cm^2/V \cdot s$ to $\sim 3000 \ cm^2/V \cdot s$ as electron dose increases to $10^{14} \ e^-/cm^2$. Since the concentration of bound surface charge is linear in beam dose, the mobility is inversely proportional to the surface charge. This indicates the role of the surface charges as charged-impurities in the long-range scattering model.

Figures 3d and 3e show the minimal conductivity ($\sigma_{min}$) and plateau width ($\Delta V_g$) at the Dirac point as a function of the reciprocal of carrier mobility, which increases with the charged-impurity density[26]. As the charged-impurity density increases, the minimal conductivity reduces while the plateau width broadens. These observations are consistent with previous reports[4,28], corroborating the role of surface charging in SiO$_2$ acting as long-range scattering centres in graphene[26,27]. More importantly, at the same impurity density (i.e. for the same value of $1/\mu$ ), the doping induced by a 30 keV beam has similar values of $\sigma_{min}$ and $\Delta V$ (blue diamond points) compared to those induced by 2 keV e-beam irradiation. It means that the carrier scattering in graphene is dominated by the near-surface impurities and independent of the deeper injected charges in the substrate.

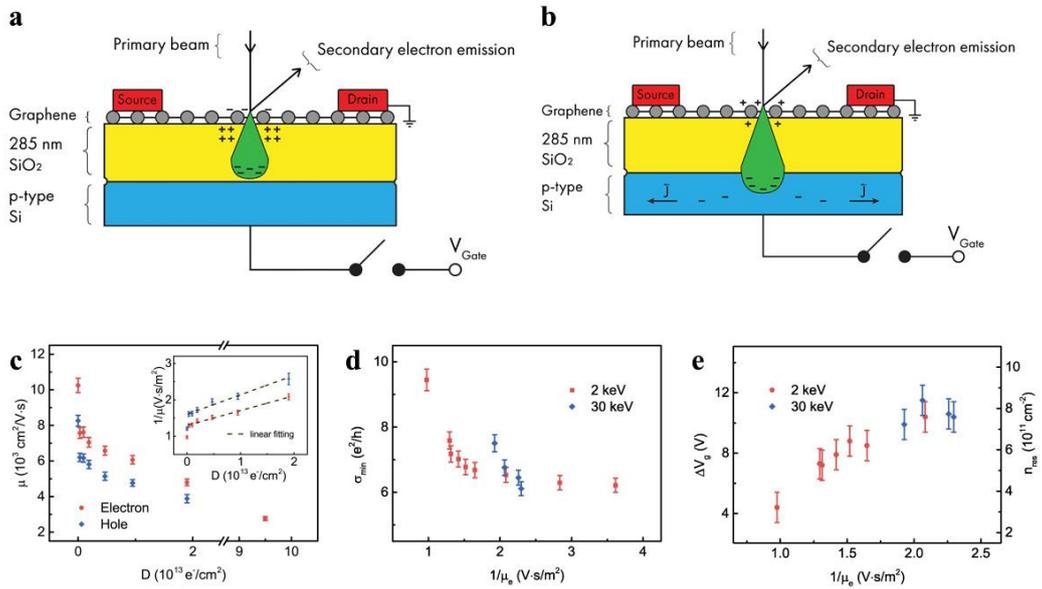

**Figure 3| Physical model and impurity scattering in graphene.** Sketches demonstrate **a** low energy (2 keV) e-beam induced n-type doping and **b** high energy (30 keV) e-beam induced p-type doping. **c,** Variation of carrier mobility as a function of electron dose. Inset:

linear relationships between inverse mobility and dose, indicating charged impurity scattering. **d**, Minimal conductivity as a function of inverse mobility. **e**, Gate plateau width (residual carrier density) as a function of inverse mobility.

**Electron beam-driven programmable graphene logic devices**

The field effect enabled by the e-beam doping is similar to the application of a back gate. The e-beam doping, however, enables logic operations of multiple types on a single chip which is impractical to achieve through back gating. This is due to the site-specific nature of the focused e-beam irradiation, and the selective doping of any region of a supported graphene flake, allowing for the formation of p-n junctions for logical operations. We first demonstrate a NOT gate on a four-terminal (input $V_{IN}$, output $V_{OUT}$, power supply $V_{DD}$, and ground GND) graphene FET. As shown in Figure 4a, region B is n-doped by a 2 keV e-beam with a dose of $10^{13}$ $e^-/cm^2$, while leaving region A in its original state. Two minima are observed in the back gate characterization curve (red curve in the inset of Fig. 4b), indicating the formation of a p-n junction. The device response changes from a constant voltage output (black dashed curve in Fig. 4b) to a logic inverter output. A high output bias (0.85 V for 1V power supply) at low input voltages ($V_{IN} < 0$) and a low output bias at high input voltages are observed (red curve in Fig. 4b). Furthermore, the logic output profile can be tuned by varying the irradiation doses, and erased using a high energy beam (Supplementary Figs. S16b and S16c), showing the tunability of the NOT gate.

This site-specific and erasable doping indicates the logic operation can be programmed by e-beam irradiation, thus allowing for multiple types of logic gates in the same device. We fabricate a graphene device with two top gates operating as logic inputs (Fig. 4c and supplementary section 11). By using this device, a NAND gate is first demonstrated when the region between V<sub>OUT</sub> and GND (blue dashed rectangle in Fig. 4c) is n-doped by a 2 keV e-beam with dose of $10^{14}\ e^-/cm^2$. The input voltages of -2 V and + 2 V correspond to binary "0" and "1" states. The output voltage V<sub>OUT</sub> is measured as a function of time with a switching interval of 30 s, while the two input states are varied across all four possible binary combinations (0,0), (0,1), (1,0) and (1,1). V<sub>OUT</sub> is seen to maintain a relatively high level between 0.85 V to 1 V for the first three input combinations, where at least one of the inputs is a "0", but decreases to the low level of 0.6 V when both inputs are set to "1". This response demonstrates the nominal NAND gate functionality for this graphene device. We then erase this doping profile with the e-beam, and instead dope the region between V<sub>DD</sub> and V<sub>OUT</sub> (red dashed rectangle in Fig. 4d). The device exhibits a different output response. A logic "0" output of ~ 0.75 V for V<sub>OUT</sub> could only be achieved when both inputs are at "0". V<sub>OUT</sub> stays at logic "1" output between 0.9 V to 1 V for all other three input combinations. Such device is an OR gate rather than the previously demonstrated NAND gate. Therefore, by modulating the doping profiles in different regions of the same graphene device, an individual graphene transistor can work as a programmable logic gate to achieve different types of switching operations.

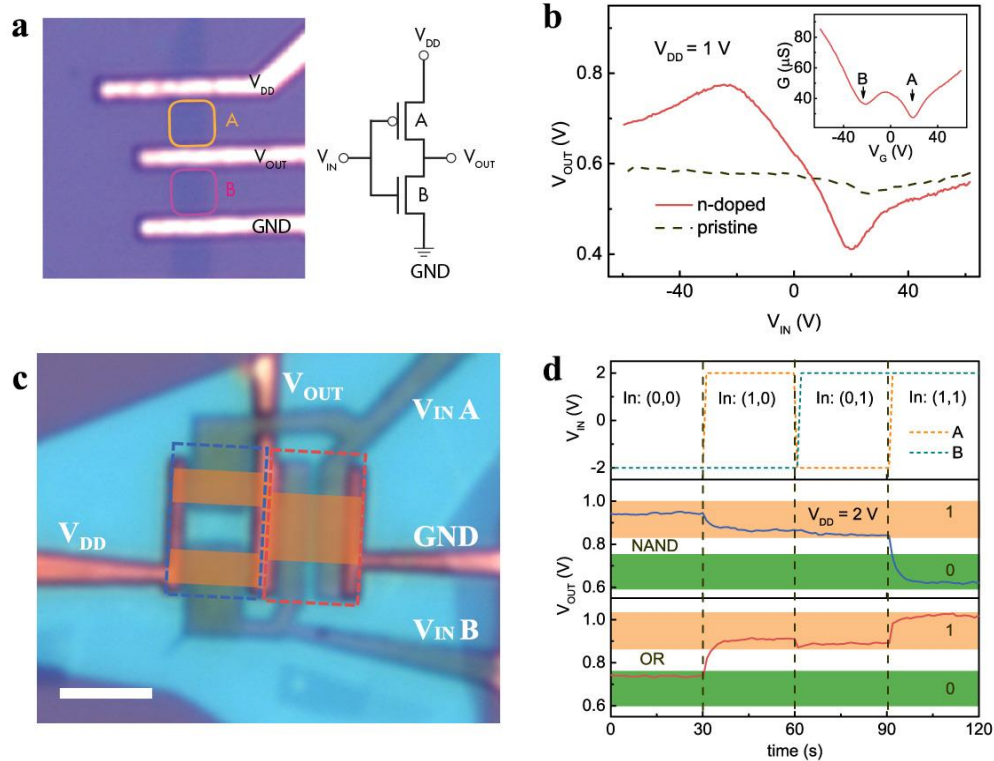

**Figure 4| Demonstration of programmable graphene logic devices. a,** optical image of a graphene logic inverter, with two distinct regions marked. Right: circuit diagram of a logic inverter. **b,** The output voltage ($V_{OUT}$) as a function of input voltage ($V_{IN}$) before and after selectively doping in region B. Power supply $V_{DD}$ = 1 V. Inset: splits of charge neutral points (marked A and B) indicating the formation of a p-n junction. **c,** optical image of a dual input graphene programmable device. The yellow shaded areas show the geometry of graphene flake. Two red and blue rectangles show the doped region in NAND and OR measurements respectively. **d**, NAND and OR logic operations in the device shown in **c** by doping different regions. A -6 V back gate is applied during the measurements to obtain a maximum difference between "0" and "1" output voltages. The scale bars in **a** and **c** are both 5 µm.

**E-beam doping in other 2D materials**

The e-beam doping can also be extended to other 2D materials. Figure 5a shows the gate response of a monolayer MoS$_2$ FET under different doses of 2 keV e-beam irradiation. MoS$_2$ exhibits intrinsic n-type doping with a FET threshold voltage (V$_{TH}$) of $\sim +20\,V$ (supplementary Fig. S18a). The e-beam irradiation monotonically shifts the gate curve to the left of this threshold as the dose increases to $10^{14}\,e^-/cm^2$, indicating the rise of the n-type doping level. We extract threshold voltages under different irradiation doses: V$_{TH}$ shifts from $\sim +20\,V$ to a saturated value of $\sim -10\,V$ after a dose of $10^{14}\,e^-/cm^2$ (see inset of Fig. 5a). Due to the band gap ($\sim$ 1.9 eV) of MoS$_2$[29], the modulation of carrier densities can tune the device conductance over several orders, allowing the creation of clearly distinguishable highly conductive "ON" and poorly conductive "OFF" states (Supplementary Fig. S18b). As demonstrated in Figure 5b, stable "ON" states with high conductivity can be written using 2 keV e-beam irradiation, and erased by 30 keV e-beam irradiation to create unstable poorly conductive "OFF" states. The writing and erasing of conducting states can be repeated for many cycles. Due to the unipolar gate response in MoS$_2$, it is difficult to obtain several types of logic operation in one MoS$_2$ device. However, a high performance programmable logic device may also be experimentally demonstrated using black phosphorus as the channel material. Few-layer black phosphorus is a semiconductor (bandgap of 0.3 eV) with an ambipolar gate response[30], which would enable both erasable p-type doping and n-type doping in the same device by e-beam irradiation. Further work will be carried out to build this programmable phosphorene logic device.

In conclusion, we have demonstrated programmable doping in graphene by localized irradiation with a focused electron beam. The derived physical model fits the experimental data and is applicable not only to graphene but also to other 2D materials, as demonstrated in the case of monolayer MoS$_2$. The proposed methodology provides high throughput in efficient, controllable and rewritable doping of on-substrate nanomaterials. Importantly, it only requires a standard SEM to implement. We hope that this experimental approach is adopted for future research on 2D nanoelectronic devices.

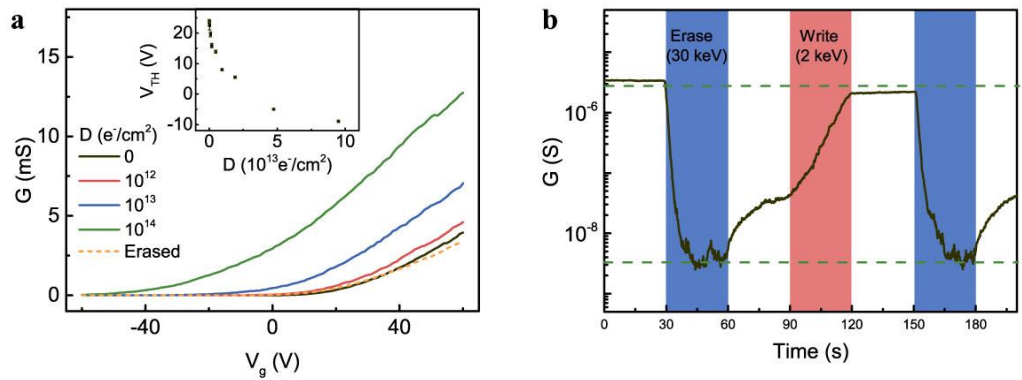

**Figure 5| E-beam doping behavior of a MoS2 FET device. a,** gate responses of a monolayer MoS$_2$ FET irradiated by a 2 keV beam with doses from $10^{12}\ e^-/cm^2$ to $10^{14}\ e^-/cm^2$ respectively. The 2 keV beam irradiation induces n-type doping and shifts the gate curve to the left, and the doping states are then erased by a 30 keV beam (orange dashed line). Inset: Change of threshold voltage as a function of e-beam dose. **b,** Switching between high and low conducting states by alternating beam energies with a time interval of 30 s. The two conducting states are marked by green dashed lines.

**Methods**

**2D FET fabrication**

Graphene and MoS$_2$ were mechanically exfoliated with adhesive tape onto a highly p-doped Si substrate capped with a 285 nm SiO$_2$ layer. The studied materials were located under an optical microscope (Olympus BX51 with a 50 × objective lens) and selected for device fabrication based on layer thickness, determined through optical contrast measurements. The flakes were first patterned to designed shaped by plasma etching using

PMMA as masks. Multiple electrical contacts were then patterned by a standard electron beam lithography (EBL) process followed by deposition of 5 nm Ti and 35 nm Au film. For the dual-input logic devices, thin boron nitride flakes (selected thickness around 20 nm determined by atomic force microscope measurements) with sizes larger than the dimensions of the FET channel were precisely placed on top of the device through a site-specific micro-stamp transfer process[31]. An additional EBL process was carried out to pattern top gate electrodes.

**In-situ electrical measurements in the SEM**

The fabricated FET devices were placed in the vacuum chamber of a Zeiss EVO SEM at a base pressure of $2 \times 10^{-5}$ mbar. The beam energy was varied from 1 keV to 30 keV. The irradiation doses were controlled by the beam current ($20 \pm 1$ pA at 2 kV and $130 \pm 1$ pA at 30 kV) and the irradiation time (controlled with Raith ELPHY Quantum software). Nano-manipulators (Imina Inc.) were used to make contacts between electrodes and the tungsten probes. The examined chips were placed on an insulator (glass slide) to isolate them from the common SEM ground. An additional detachable cable was attached to the Si layer as the back gate control. Reduced scan was used to contact the probes with electrode pads to avoid any e-beam exposure to the graphene flake.

The two-terminal electrical measurements were carried out by using a dual channel source-measurement unit (Keysight B2912A). One channel worked as a drain-source power supply (constant voltage of 0.1 V) and the other channel worked as back gate control (typical gate voltage ranging from – 60V to + 60V). Four-terminal electrical characterisation was carried out by the combination of a lock-in amplifier (Stanford SR 830) as a drain-source power supply (constant AC current of 100 nA with low frequency of 17.777 Hz) and Keysight B2912A as a back gate control. The single-input logic inverter measurements were carried out using two channels of Keysight B2912A as power supply ($V_{DD}$) and logic input ($V_{IN}$) respectively, while a Keysight 34420A nanovoltmeter was applied to measure the output voltage $V_{OUT}$. In the dual-input programmable logic devices, the input signals used DC output from the lock-in amplifier (ranging from -10.5 V to + 10.5 V).

**Acknowledgements**

We would like to thank the staff at the Advanced Microscopy Laboratory (AML), CRANN, Trinity College Dublin. We would like to acknowledge support from the following funding bodies: the Science Foundation Ireland [grant No: 11/PI/1105, No. 07/SK/I1220a, and 08/CE/I1432] and the Irish Research Council [grant No: GOIPG/2014/972].

**Author Contributions**

Y.Z., J.J. and D.K. prepared the devices. Y.Z., J.J and D.K. carried out the in-situ electrical measurements. Y.Z., J.J., D.K. and H.Z analyzed the data. Y.Z., J.J., D.K., Y.C., D.Y. and H.Z. wrote the paper. H.Z. supervised the project and led the overall effort. All authors interpreted and discussed the experimental results and edited the manuscripts. Y.Z., J.J and D.K. contributed equally to this work.

**Additional information**

**Supplementary information:** The supplementary information can be found in the attachment.

**Competing financial interests:** The authors declare no competing financial interests.